\documentclass{aa}  

\usepackage{amssymb}
\usepackage{amsmath}
\usepackage{graphicx}
\usepackage{txfonts}
\usepackage{booktabs}
\usepackage{textcomp}
\usepackage{color}
\usepackage{lscape}
\usepackage{longtable}
\usepackage{xcolor}
\usepackage{multirow}
\bibpunct{(}{)}{;}{a}{}{,}

\newcommand{\HXMT}{\emph{Insight--\/}HXMT}
\newcommand{\FRB}{\object{FRB\,180916.J0158$+$65}}

\begin{document} 

  \title{Constraining the transient high-energy activity of \FRB\ with \HXMT\ followup observations}
\titlerunning{X/$\gamma$--ray followup observations during active phase of Feb 4-7, 2020}
\authorrunning{C.~Guidorzi et al.}
 
   \author{C.~Guidorzi\thanks{guidorzi@fe.infn.it}\inst{1,2,3} \and M.~Orlandini\inst{3} \and F.~Frontera\inst{1,3}
   \and L.~Nicastro\inst{3} \and S.L.~Xiong\inst{4} \and J.Y.~Liao\inst{4} \and G.~Li\inst{4} \and S.N.~Zhang\inst{4,5} \and L.~Amati\inst{3} \and E.~Virgilli\inst{3} \and S.~Zhang\inst{4} \and Q.C.~Bu\inst{4} \and C.~Cai\inst{4,5} \and X.L.~Cao\inst{4} \and Z.~Chang\inst{4} \and L.~Chen\inst{6} \and T.X.~Chen\inst{4} \and Y.~Chen\inst{4} \and Y.P.~Chen\inst{4} \and W.W.~Cui\inst{4} \and Y.Y.~Du\inst{4} \and G.H.~Gao\inst{4,5} \and H.~Gao\inst{4,5} \and M.~Gao\inst{4} \and M.Y.~Ge\inst{4} \and Y.D.~Gu\inst{4} \and J.~Guan\inst{4} \and C.C.~Guo\inst{4,5} \and D.W.~Han\inst{4} \and Y.~Huang\inst{4} \and J.~Huo\inst{4} \and S.M.~Jia\inst{4} \and W.C.~Jiang\inst{4} \and J.~Jin\inst{4} \and L.D.~Kong\inst{4,5} \and B.~Li\inst{4} \and C.K.~Li\inst{4} \and T.P.~Li\inst{4,5,7} \and W.~Li\inst{4} \and X.~Li\inst{4} \and X.B.~Li\inst{4} \and X.F.~Li\inst{4} \and Z.W.~Li\inst{4} \and X.H.~Liang\inst{4} \and B.S.~Liu\inst{4} \and C.Z.~Liu\inst{4} \and H.X.~Liu\inst{4,5} \and H.W.~Liu\inst{4} \and X.J.~Liu\inst{4} \and F.J.~Lu\inst{4} \and X.F.~Lu\inst{4} \and Q.~Luo\inst{4} \and T.~Luo\inst{4} \and R.C.~Ma\inst{4,5} \and X.~Ma\inst{4} \and B.~Meng\inst{4} \and Y.~Nang\inst{4,5} \and J.Y.~Nie\inst{4} \and G.~Ou\inst{8} \and J.L~Qu\inst{4} \and X.Q.~Ren\inst{4,5} \and N.~Sai\inst{4,5} \and L.M.~Song\inst{4,5} \and X.Y.~Song\inst{4} \and L.~Sun\inst{4} \and Y.~Tan\inst{4} \and L.~Tao\inst{4} \and Y.L.~Tuo\inst{4,5} \and C.Wang\inst{9,5} \and L.J.~Wang\inst{4} \and P.J.~Wang\inst{4,5} \and W.S.~Wang\inst{8} \and Y.S.~Wang\inst{4} \and X.Y.~Wen\inst{4} \and B.Y.~Wu\inst{4,5} \and B.B.~Wu\inst{4} \and M.~Wu\inst{4} \and G.C.~Xiao\inst{4,5} \and S.~Xiao\inst{4,5} \and Y.P.~Xu\inst{4,5} \and R.~Yang\inst{10} \and S.~Yang\inst{4} \and Y.J.~Yang\inst{4} \and Q.B.~Yi\inst{4,11} \and Q.Q.~Yin\inst{4} \and Y.~You\inst{4,5} \and F.~Zhang\inst{4} \and H.M.~Zhang\inst{8} \and J.~Zhang\inst{4} \and P.~Zhang\inst{4} \and W.C.~Zhang\inst{4} \and W.~Zhang\inst{4,5} \and Y.F.~Zhang\inst{4} \and Y.H.~Zhang\inst{4,5} \and H.S.~Zhao\inst{4} \and X.F.~Zhao\inst{4,5} \and S.J.~Zheng\inst{4} \and Y.G.~Zheng\inst{4,10} \and D.K.~Zhou\inst{4,5}
}
\institute{Department of Physics and Earth Science, University of Ferrara, via Saragat 1, I--44122, Ferrara, Italy
\and INFN -- Sezione di Ferrara, via Saragat 1, I--44122, Ferrara, Italy
\and INAF -- Osservatorio di Astrofisica e Scienza dello Spazio di Bologna, Via Piero Gobetti 101, I-40129 Bologna, Italy
\and Key Laboratory of Particle Astrophysics, Institute of High Energy Physics, Chinese Academy of Sciences, 19B Yuquan Road, Beijing 100049, People’s Republic of China
\and University of Chinese Academy of Sciences, Chinese Academy of Sciences, Beijing 100049, China
\and Department of Astronomy, Beijing Normal University, Beijing 100088, People’s Republic of China
\and Department of Astronomy, Tsinghua University, Beijing 100084, People’s Republic of China
\and Computing Division, Institute of High Energy Physics, Chinese Academy of Sciences, 19B Yuquan Road, Beijing 100049,People’s Republic of China
\and Key Laboratory of Space Astronomy and Technology, National Astronomical Observatories, Chinese Academy of Sciences, Beijing 100012,People’s Republic of China 
\and College of physics Sciences and Technology, Hebei University,
No. 180 Wusi Dong Road, Lian Chi District, Baoding City, Hebei Province, 071002 China
\and School of Physics and Optoelectronics, Xiangtan University, Yuhu District, Xiangtan, Hunan, 411105, China
}
\date{}

\abstract
   {A link between magnetars and fast radio burst (FRB) sources has finally been established. In this context, one of the open issues is whether/which sources of extragalactic FRBs exhibit X/$\gamma$--ray outbursts and whether it is correlated with radio activity.}
   {We aim to constrain possible X/$\gamma$--ray burst activity from one of the nearest extragalactic FRB sources currently known over a broad energy range, by looking for bursts over a range of timescales and energies that are compatible with being powerful flares from extragalactic magnetars.}
   {We followed up the as-yet nearest extragalactic FRB source at a mere 149~Mpc distance, the periodic repeater \FRB, during the active phase on February 4--7, 2020, with the {\em Insight}--Hard X--ray Modulation Telescope (\HXMT). Taking advantage of the combination of broad band, large effective area, and several independent detectors available, we searched for bursts over a set of timescales from 1~ms to $1.024$~s with a sensitive algorithm, that had previously been characterised and optimised. Moreover, through simulations we studied the sensitivity of our technique in the released energy-duration phase space for a set of synthetic flares and assuming different energy spectra.}
   {We constrain the possible occurrence of flares in the 1-100~keV energy band to $E<10^{46}$~erg for durations $\Delta\,t<0.1$~s over several tens of ks exposure.}
   {We can rule out the occurrence of giant flares similar to the ones that were observed in the few cases of Galactic magnetars. The absence of reported radio activity during our observations does not allow us to make any statements on the possible simultaneous high-energy emission.}
\maketitle
%
\newcommand{\commcg}[1]{\textcolor{red}{Cristiano: #1}}
\newcommand{\commmo}[1]{\textcolor{olive}{Mauro: #1}}
\newcommand{\commff}[1]{\textcolor{cyan}{Filippo: #1}}
\newcommand{\commln}[1]{\textcolor{blue}{Luciano: #1}}
\newcommand{\commla}[1]{\textcolor{orange}{Lorenzo: #1}}

\section{Introduction}
Fast radio bursts (FRBs) are a new class of ms-long radio flashes of unknown 
extragalactic origin, for which the host galaxy identification and redshift determination have become feasible in several cases in recent years \citep{Tendulkar17,Bannister19,Prochaska19,Ravi19b,Marcote20,Macquart20}.
The combination of short duration, specific luminosity ($\lesssim 10^{34}$~erg\,s$^{-1}$\,Hz$^{-1}$), and brightness temperature ($T_b\gtrsim 10^{35}$~K) suggests that the progenitor is a compact source emitting through a coherent process (see \citealt{CordesChatterjee19,Petroff19_rev} for recent reviews).

Especially after the discovery of repeating FRB sources \citep{Spitler16,CHIME19b,CHIME19c,Kumar19,Fonseca20} magnetars caught further attention as one of the most promising candidates \citep{Popov10,Lyubarsky14,Beloborodov17,Metzger19}. From the theoretical side, connections with sources of gamma--ray bursts (GRBs) are not ruled out (see \citealt{Platts19} for a review of theoretical models), as young ms-magnetars could be the endpoint of GRB progenitors \citep{Usov92,Thompson94,Bucciantini07,Metzger11}, and cataclysmic models cannot be ruled out as long as one-off FRBs are observed. Nonetheless, observations carried out so far seem to exclude a systematic link between FRBs and standard cosmological GRBs \citep{Tendulkar16,Guidorzi19,Guidorzi20,Martone19a,Cunningham19,Anumarlapudi20}.
The possibility, suggested by \citet{RaviLasky14}, that a FRB could result from the final collapse of a newborn supramassive neutron star some $10$ to $10^4$ s after a binary neutron star merger, which would be signalled by a short GRB, found no confirmation through the search for FRB counterparts in the case of four promptly localised short GRBs \citep{Bouwhuis20}.

The extreme magnetic field ($B\sim 10^{14}$--$10^{15}$~G) of magnetars is thought to power their high-energy emission, which is characterised by periods of quiescence, interspersed with active intervals described by sporadic short X--ray bursts (typical duration of $\sim 0.1$~s) with luminosities in the range $10^{36}$--$10^{43}$~erg\,s$^{-1}$, and rarely by giant flares (GFs). These consist of an initial short spike with a peak luminosity in the range $10^{44}$--$10^{47}$~erg/s, followed by a several-hundred-second-long fainter tail modulated by the star's spin period. Only three GFs from magnetars, two in the Galaxy and one in the Large Magellanic Cloud, have been observed so far, although a few extragalactic candidates have also been reported \citep{Frederiks07,Mazets08,Svinkin20,Frederiks20,Yang20}. See \citet{Turolla15,Mereghetti15,KaspiBeloborodov17} for reviews on magnetars.

For most FRBs the possibility of a simultaneous magnetar GF could not be discarded, mainly due to the limited sensitivity of past or currently flying $\gamma$--ray detectors, combined with FRB distances (e.g., \citealt{Martone19a,Guidorzi19,Guidorzi20}). Yet, ignoring possible different beaming factors between radio and high-energy emission, a one-to-one correspondence between FRBs and magnetar GFs has to be excluded because of the meaningful lack of a radio detection associated with the most luminous Galactic GF yet observed \citep{Tendulkar16}.

A turning point recently came with the discovery of Galactic FRB\,200428 simultaneously with a hard X--ray burst from lately reactivated magnetar SGR\,1935+2154 \citep{LiHXMT20,Mereghetti20,Ridnaia20}: as seen with the Canadian Hydrogen Intensity Mapping Experiment (CHIME; \citealt{CHIME19b}), the FRB consists of two peaks 30-ms apart, with an energy of $3\times10^{34}$~erg and peak luminosity of $7\times10^{36}$~erg\,s$^{-1}$ in the 400--800-MHz band \citep{CHIME20b}. It was also detected with the Survey for Transient Astronomical Radio Emission 2 (STARE2; \citealt{BochenekSTARE2}) in the 1281--1468~MHz band with a burst energy of $2\times10^{35}$~erg and a luminosity of $4\times 10^{38}$~erg\,s$^{-1}$ \citep{Bochenek20}. This FRB is $\sim40$ times less energetic than the least energetic extragalactic one so far measured. The X/$\gamma$-ray counterpart also consists of two main peaks temporally coincident with their radio analogues, once the delay expected from the dispersion measure is accounted for. Interestingly, while the released energy is typical of magnetar bursts, this event exhibits an unusually structured, slowly rising time profile compared with that of a typical short burst. In addition, its spectrum, unusually hard as well, is fitted with a cutoff power-law with photon index between $0.4$ and $1.6$ and cutoff energy in the range $65$--$85$~keV, corresponding to a released energy of $1\times10^{40}$~erg and peak luminosity of $1\times10^{41}$~erg\,s$^{-1}$ \citep{LiHXMT20,Ridnaia20}. This FRB followed an intense hard X--ray burst activity which had culminated with a burst forest on April 27, 2020 \citep{Palmer20,Younes20}.
On the one hand, this event finally provides direct evidence that magnetars can originate FRBs along with hard X--ray bursts; on the other hand, the lack of radio counterparts to many other hard X--ray bursts from the same source, with upper limits $10^8$ times fainter than FRB\,200428, shows the rarity of this kind of joint emission \citep{Lin20}.

The discovery of \FRB, the as-yet nearest extragalactic FRB source with measured redshift, at a luminosity distance of $149.0\pm 0.9$~Mpc \citep{Marcote20} along with its being a repeater, made it a desirable target for multi-wavelength  surveys. The subsequent discovery of a periodic modulation in its radio burst activity with $P=16.35\pm0.15$~days, with a FRB rate of up to $\sim 1$~hr$^{-1}$ for a $\pm2.7$-day window around peak \citep{CHIME20a}, opened up to planning multiwavelength campaigns around the expected peaks. In a 33-ks {\em Chandra X-ray Observatory} observation which covered one FRB detected with CHIME, \citet{Scholz20} detected no X--ray source, with upper limits on the released energy of $1.6\times10^{45}$~erg and $4\times10^{45}$~erg at the FRB time and at any time, respectively, in the $0.5$--$10$-keV energy band. They also derived an upper limit of $6\times10^{46}$~erg in the $10$--$100$~keV band for 12 bursts from \FRB\ that were visible with {\em Fermi}/Gamma-ray Burst Monitor (GBM; \citealt{Meegan09}).
{\em XMM-Newton} detected no source down to $E<10^{45}$~erg in the $0.3$--$10$-keV energy band at the times of three radio bursts, that were discovered at 328~MHz with the Sardinia Radio Telescope \citep{Pilia20}.
Comparable upper limits of $3\times10^{46}$~erg on the energy released in the optical band simultaneous with FRBs from \FRB\ were also derived in a statistical framework using survey data of the Zwicky Transient Facility \citep{Andreoni20}.

The Hard X--ray Modulation Telescope (HXMT), named ``Insight'' after launch  on June 15, 2017, is the first Chinese X--ray astronomy satellite  \citep{Li07,Zhang20_HXMT}. It carries on board three main  instruments: the Low Energy X--ray telescope (LE; 1--15~keV;  \citealt{Chen20_HXMT}), the Medium Energy X--ray telescope (ME; 5--30~keV; 
\citealt{Cao20_HXMT}), and the High Energy X--ray telescope (HE; \citealt{Liu20_HXMT}). The HE consists of 18 NaI/CsI detectors covering the 20--250~keV energy band for pointing observations. Moreover, it also works as an all-sky monitor in the $0.2$--3~MeV energy range. The unique combination of a very large geometric area ($\sim5100$\,cm$^{2}$) and of 
continuous event tagging with timing accuracy $<10\,\mu$s, was already exploited in the search for possible $\gamma$--ray counterparts to a sample of 39~FRBs down to ms or sub-ms scales in the keV--MeV energy range. As a result, the association with cosmological GRBs was excluded on a systematic basis \citep[hereafter G20]{Guidorzi20}.

In this work we report the results of followup observations of \FRB\ that were carried out with \HXMT\ around one of the expected peaks of radio activity, from February 4 to 7, 2020, for which no observations have been reported to date.

Section~\ref{sec:dataset} describes data set and reduction, whereas the analysis is  reported in Section~\ref{sec:data_an}. Results are in reported in Section~\ref{sec:res} and discussed in Section~\ref{sec:disc}. We conclude in Section~\ref{sec:conc}.



\section{Data set}
\label{sec:dataset}
\HXMT\ observed \FRB\ from 2020-02-04 12:40:19 to 2020-02-07 07:39:44 UT as a Target of Opportunity observation, requested in correspondence of a predicted maximum from radio observations.  The detailed log of the \HXMT\ observation is listed in Table~\ref{tab:log}.  The different net exposures of the instruments are due to the filtering criteria adopted for the generation of the good time intervals.
\begin{table*}
\caption{Log of the \HXMT\ observation of \FRB. The observation lasted 241142~s, from 2020-02-04 12:40:19 to 2020-02-07 07:39:44 UT.}
\label{tab:log}
\centering\begin{tabular}{ccrrr}
\hline\hline
Inst & Energy & Net Exp$^{\rm (a)}$ & Net Exp$^{\rm (b)}$ & Net Count Rate$^{\rm (c)}$ \\
& (keV)  & (ks) & (ks)& (c/s) \\
\hline
LE &  1--10 & $29.7$ & $29.2$ & $0.452 \pm 0.022$ \\
ME & 10--30 & $67.3$ & $67.1$ & $-0.015 \pm 0.027$ \\
HE & 25--80 & $47.0$ & $46.5$ & $-2.516 \pm 0.374$ \\
LE \& ME \& HE & 1--80 & -- &  $19.2$ & --\\
LE \& ME$^{\rm (d)}$  & 1--30 & -- &  $9.9$ & --\\
ME \& HE$^{\rm (d)}$  & 25--80 & --  &  $23.9$ & --\\
HE alone$^{\rm (e)}$ & 25--80 & -- & $3.3$ & --\\
\hline
\end{tabular}
\begin{list}{}{}
\item[$^{\rm (a)}$]{Net exposure resulted from the the standard filtering described in the text.}
\item[$^{\rm (b)}$]{Net exposure resulted after applying the background interpolation procedure.}
\item[$^{\rm (c)}$]{Derived with HXMTDAS tasks (see Section~\ref{sec:dataset}).}
\item[$^{\rm (d)}$]{Filtered data are available only for two instruments.}
\item[$^{\rm (e)}$]{Neither LE nor ME have simultaneous filtered data.}
\end{list}
\end{table*}

For our analysis we used the software package HXMTDAS version 2.02.1\footnote{\url{http://enghxmt.ihep.ac.cn/software.jhtml}}. The screening of the raw events was performed by means of the the \texttt{legtigen, megtigen,} and \texttt{hegtigen} tasks\footnote{\url{http://enghxmt.ihep.ac.cn/SoftDoc.jhtml}}. The standard filtering criteria were adopted, namely: the Earth elevation angle $\mathrm{ELV}>10\deg$; the cutoff rigidity $\mathrm{COR}>8$~GeV; the pointing offset angle $\mathrm{ANG\_DIST}<0.04\deg$. We also excluded data taken close to the South Atlantic Anomaly (SAA) by selecting T\_SAA and TN\_SAA both greater than 300~sec. For the LE instrument, we selected only data for which the Bright Earth Angle was greater than $30\deg$.

From the cleaned event files we then extracted the light curves with 1~ms time resolution. For the HE
instrument we also extracted the light curves for each HE unit, in order to improve the efficiency of the Multi-Detector-Search algorithm to the LE+ME+HE light curves.

Because of the uncertainties in the background evaluation (as evident by the net count rates listed in Table~\ref{tab:log}), performed by the tasks \texttt{lebkgmap} \citep{LEBCKG}, \texttt{mebkgmap} \citep{MEBCKG}, and \texttt{hebkgmap} \citep{HEBCKG}, the background was evaluated by applying a procedure that interpolates the background starting from a 10-s binned light curve, applying polynomials with increasing order to individual orbits up to the point where both $\chi^2$s and runs tests (2-tails) had a P-value $>0.01$, so as to avoid both under- and over-fitting. The time bins for which this procedure could  not come up with the required P-values were discarded and the resulting net exposure is also reported in Table~\ref{tab:log}. Consequently, this procedure can only detect relatively short bursts, whereas a possible relatively faint, constant source, or varying over timescales $>10$~s cannot be detected.
Table~\ref{tab:log} also reports the net exposure for each combination of instruments, whose data are available simultaneously: only the LE and HE combination without ME does not appear to have a significant exposure. We included the time intervals for which only HE data are available, because the multiplicity of its independent detectors still enables an effective search for transients. Concerning the HE, in the following we ignored the blocked collimator detector, which was devised to measure the local background of HE \citep{Liu20_HXMT}, therefore using the data from the remaining 17 detectors.
Hereafter, only the filtered time bins for each instrument are considered in the analysis.

There is no FRB reported during our observations: in particular, on 2020-02-04 CHIME reported four bursts, whose last one was at 01:17:21.37~UT, so more than 11 hours before the beginning of \HXMT\ observations. The next FRB reported by CHIME from this source was 15 days later \footnote{\url{https://www.chime-frb.ca/repeaters/180916.J0158+65}.}.
Assuming the period of $16.35$~days, the expected peak of radio activity considered by us was at 2020-02-05 04:55~UT: the \HXMT\ observing window spans the time interval from $-0.7$ to $2.1$~days around it, so completely within the $\pm 2.7$-day interval characterised by the expected peak burst rate of $1.0\pm0.5$~hr$^{-1}$ \citep{CHIME20a}. The total net exposure used in the present work is $0.65$~days, which corresponds to 23\% of the overall observing window.

\section{Data analysis}
\label{sec:data_an}
A transient increase of the count rate of the detectors can be caused by two different kinds of phenomena: (i) an electromagnetic wave associated with a transient event, whose photons interact with a number of detectors; (ii) high-energy charged particles interacting with individual detectors. The main distinctive property of the e.m. wave is a common spectral and temporal evolution as recorded by the different detectors, whereas the particle-induced event deposits its energy in one or a few detectors and in an uncorrelated way, resulting in spikes in the count rates significantly in excess of what is expected from counting statistics. In particular, when dim astrophysical transients and short integration times are considered, the very few expected counts can be more easily confused with particle spikes. Consequently, searching for simultaneous excesses over a number of detectors is the most effective way for discriminating them.

To this aim, G20 developed the so-called multi-detector search (hereafter MDS) method that exploited the segmented nature of the \HXMT/HE instrument to search for transient candidates possibly associated with FRBs. While in that case only CsI events were considered (being transparent to the collimators), here the data to be analysed are from a pointed observation and mainly differ in two aspects: 1) concerning the HE, NaI instead of CsI events are considered; 2) the data of the other two instruments operating in the corresponding softer energy bands are included. In the light of this, we had to tweak and adapt the original MDS algorithms as described below. In order to find the optimal compromise between sensitivity and false positive rate, we preliminarily characterised the background statistical properties for each detector.

\subsection{Background statistical properties}
\label{sec:statnoise}
Prior to investigating the nature of possible candidates, we assumed that their signal does not affect the overall count distribution, since the great majority of the recorded counts is assumed to be background. For each of the 19 detectors (LE, ME, and 17 HE-NaI units) we accumulated the overall 1-ms count distribution. In order to test whether this is compatible with being a statistical realisation of a variable Poisson process, whose expected value for each time bin is given by the locally estimated background, we simulated 100 realisations for each bin. For each detector we thus ended up with a distribution of expected counts having 100 times as many bins as the corresponding real one.

We then compared each of the 19 real count distributions with their corresponding synthetic ones. As a result, the total recorded counts are slightly, but significantly in excess of pure Poisson noise by the following amounts: $1.2$\%, $0.6$\%, and $0.5$\% for the LE, ME, and average HE, respectively). These are caused by the occasional presence of spikes that are visible in individual detectors or, in any case, incompatible with the signal that is expected from a plane wave. We also found that a possible way to reject most of them is increasing the lower threshold on the photon energy, at least for the HE units.

Although this excess component accounts for $\lesssim 1$\% of the total background variance, it can affect the estimate of the statistical significance of peaks in the light curves, especially at short ($\sim$~few ms) integration times. It must be therefore taken into account in calculating the expected false positive rate. More details are reported in Appendix~\ref{sec:app_A}.

\subsection{Multi-detector search}
\label{sec:MDS}
The diversity of the three instruments and energy bands, coupled with the different combinations of available data shown in Table~\ref{tab:log}, forced us to conceive a set of three complementary trigger criteria, that address several alternative cases, in which a candidate can be found in principle. For each case, the philosophy is the same as the one of the MDS conceived in G20: for a given criterion, the threshold that must be exceeded by the counts for a generic bin, depends on (a) the local interpolated background; (b) the integration time; (c) the minimum number of detectors to be triggered simultaneously, so as to end up with a desired combined probability. In addition, in the present work the threshold must also depend on the kind of detector as well as on the data available at any given time bin. Following these guidelines, for each case we have come up with a set of thresholds, expressed in units of Gaussian $\sigma$'s following the same convention as in G20. Overall, a candidate must fulfil at least one trigger criterion. A detailed description is reported in Appendix~\ref{sec:app_B}.

Because of the presence of a small, but significant extra-Poissonian variance in the background counts (Sect.~\ref{sec:statnoise}), we had to ensure that the false positive rate was not underestimated, or, equivalently, that the confidence level of any possible candidate is not overestimated. Therefore, we further calibrated the thresholds by running the MDS on 100 synthetic samples, that were obtained by shuffling all the 1-ms bins along with their associated counts and expected background counts for any detector, independently of each other. This procedure preserves the properties of the count distribution for each detector, while at the same time it offers a way for calculating the probability for any possible combination of simultaneous excesses in different detectors. In this way we ended up with a robust procedure for estimating the related multivariate probability distribution, having relaxed any assumption on the nature of the statistical noise of any individual detector. More details can be found in Appendix~\ref{sec:app_B}. Not only can the MDS be used for other FRB sources that will be targeted by \HXMT, but it may also help identify bursts from other sources not necessarily related to FRBs, such as weak short GRBs possibly associated with gravitational wave sources.

\section{Results}
\label{sec:res}
Table~\ref{tab:FPrate} reports the results of the number of candidates as a function of the integration time along with the corresponding number of expected false positives, which already accounts for the multi-trials related to the total number of time bins that were screened. We found only one candidate from the screening of 10-ms time bins, centred at 2020-02-06 23:54:55.793 UT: with reference to the three MDS criteria (Appendix~\ref{sec:app_B}), this event triggered criterion 2, that is, both LE and ME exceeded their thresholds, while only one of the HE units did. Should this be real, it would be a relatively spectrally soft event. The number of expected false positives for 10-ms integration time is $0.10$, so that the chance probability of having at least one fake candidate is $9.5$\%. More correctly, when the same probability is calculated taking into account the trials related to all the explored integration times together, the total number of expected false positives rises to $1.06$, i.e. fully consistent with the only candidate.
In order to better evaluate its nature, we also inspected its counts in both LE and ME, and found that they were just above the respective thresholds. 
We found no simultaneous events reported by {\em Fermi}/GBM, {\em INTEGRAL} SPI-ACS, {\em Swift}/BAT, and Konus/WIND. The search for coincident subthreshold triggers in the case of {\em Fermi}/GBM\footnote{\url{https://gcn.gsfc.nasa.gov/fermi_gbm_subthresh_archive.html}} and of {\em Swift}/BAT\footnote{\url{https://gcn.gsfc.nasa.gov/gcn/swift_sub_sub_threshold.html}} gave no results either. 
Overall, we conclude that, based on \HXMT\ data alone, we cannot reject the possibility that the candidate is not astrophysical.
\begin{table}
\caption{Number of candidates and expected false positives as a function of the integration time.}
\label{tab:FPrate}
\centering\begin{tabular}{rcc}
\hline\hline
$\Delta\,t^{\rm (a)}$ & $N_{\rm exp}^{\rm (b)}$ & $N_{\rm cand}^{\rm (c)}$ \\
(ms) & & \\
\hline
   1 & $0.42$ & 0\\
   4 & $0.03$ & 0\\
  10 & $0.10$ & 1\\
  64 & $0.07$ & 0\\
 256 & $0.26$ & 0\\
1024 & $0.18$ & 0\\
\hline
\end{tabular}
\begin{list}{}{}
\item[$^{\rm (a)}$]{Integration time of a single bin.}
\item[$^{\rm (b)}$]{Expected number of false positives.}
\item[$^{\rm (c)}$]{Number of candidates.}
\end{list}
\end{table}
%

\subsection{Upper limits and technique sensitivity}
\label{sec:ULsens}
Given the lack of a confident detection of transient candidates with durations in the range $10^{-3}$--$1$~s, we derived corresponding upper limits on fluence, as a function of duration, by assuming three different energy spectra: a non-thermal power-law with photon index $\Gamma=2$, which is often found to adequately describe the photon spectrum of high-energy transient events, and an optically thin thermal bremsstrahlung (hereafter, {\sc ottb}) $dN/dE\propto E^{-\Gamma}\,\exp{(-E/E_0)}$, with index $\Gamma=1$ and two different values for the cutoff energy $E_0$: $200$ and $50$~keV. Concerning the cutoff power-law, we opted for an {\sc ottb}, because it was adopted to fit the initial spikes of the few Galactic magnetar giant flares \citep{Mazets99,Feroci99,Hurley99,Hurley05,Palmer05,Frederiks07b}, some extragalactic magnetar giant flare candidates \citep{Frederiks07,Mazets08,Frederiks20}, as well as the hard X--ray burst from SGR\,1935+2154 associated with FRB\,200428, when $\Gamma$ is left free to vary from 1 \citep{LiHXMT20,Ridnaia20,Mereghetti20}.

\begin{figure*}
\centering
\includegraphics[width=\linewidth]{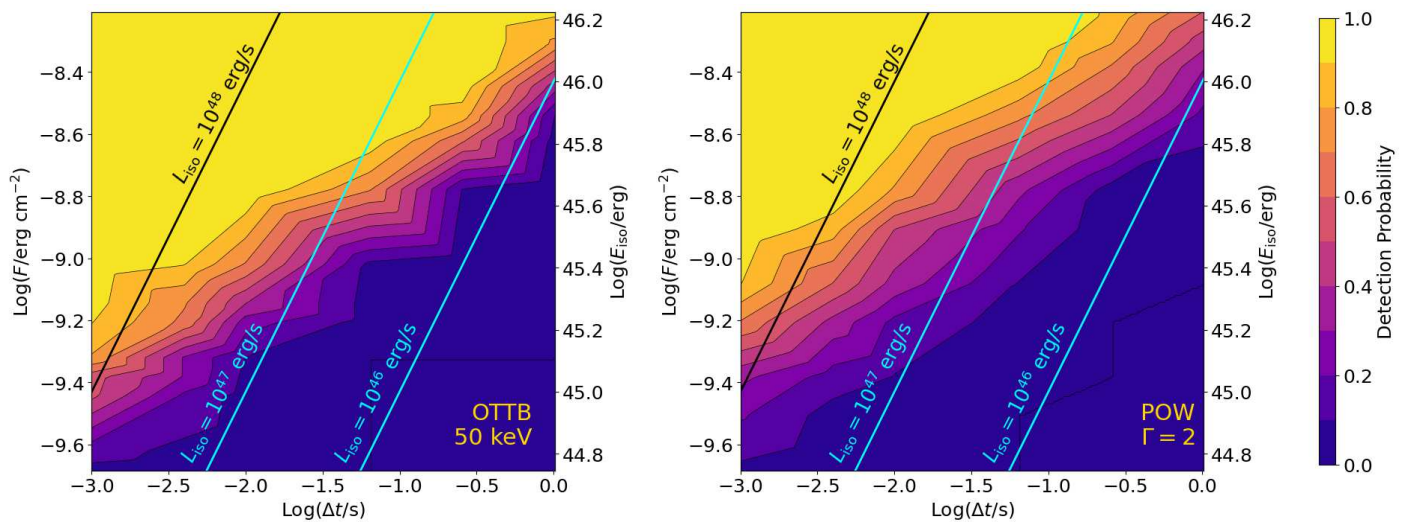}
\caption{Detection probability for a flare as a function of duration $\Delta\,t$, fluence $F$ (left-hand $y$ axis) and isotropic--equivalent released energy $E_{\rm iso}$ (right-hand $y$ axis) in the 1--100~keV energy band for two different energy spectra: an {\sc ottb} with $kT=50$~keV ({\em left}), and a power-law with photon index $\Gamma=2$ ({\em right}). The different solid lines correspond to constant luminosity values.}
\label{fig:sensitivity}
\end{figure*}

We then characterised the sensitivity of the MDS as follows: we defined a grid of points in the $F$--$\Delta\,t$ plane, where $F$ is the 1--100-keV fluence and $\Delta\,t$ is the duration of a hypothetical transient. For each spectral model and for each point of this grid we simulated 200 synthetic transients, that were added to the real counts at a given set of times uniformly distributed along the entire observing window, and counted how many of them were identified by the MDS. The results for the {\sc ottb} with $E_0=50$~keV and for the power-law are shown in the contour plots of Figure~\ref{fig:sensitivity}. The results for the {\sc ottb} with $E_0=200$~keV are omitted for the sake of clarity, because they are intermediate between the other two.
%
%
We also calculated the corresponding isotropic-equivalent released energy in the same energy band, $E_{\rm iso}$ (right-hand vertical axes in Fig.~\ref{fig:sensitivity}) at the distance of \FRB\ and, in addition, we show the lines corresponding to a constant luminosity for three different values: $10^{46}$, $10^{47}$, and $10^{48}$~erg/s. Looking at the regions with 90\% probability for a transient to be detected, for events as short as a few ms, the minimum detectable energies are $\sim 10^{45.6}= 4\times10^{45}$~erg. In terms of luminosity, the corresponding minimum values are a few $\times10^{48}$~erg/s.
Considering longer transients, up to $\sim 0.1$~s, the minimum detectable energy and luminosity values become respectively $\sim10^{46}$~erg and $10^{47}$~erg/s in the worst case.

\subsection{Any radio bursts during \HXMT\ observations?}
\label{sec:simultFRB}
Although no radio burst has been reported to date during the \HXMT\ observing window, it is worth estimating the probability that \FRB\ gave no FRBs. Ignoring the complex dependence on frequency \citep{CHIME20a,Pilia20,Chawla20}, here we focus on the homogeneous sample detected with CHIME.

Considering the $\pm0.9$-d interval centred on the peak of radio activity, which has a burst rate of $1.8_{-0.8}^{+1.3}$~hr$^{-1}$ \citep{CHIME20a}, the net exposure of \HXMT\ is $\sim 8.5$~hr. The probability of no FRBs while \HXMT\ was observing is $2\times10^{-4}$ at most. This would suggest that \HXMT\ observations covered one FRB at least, and more probably a few of them (the probability of $\le 2$ FRBs is $<1$\%). This holds true as long as a constant burst rate is assumed for the same window around all peaks. While it was already shown that the burst rate changes significantly for windows with different durations around the peak times \citep{CHIME20a}, nothing is said on whether the constant rate assumption for a given window over different peak times is compatible with observations.

We therefore tested this possibility by taking the CHIME exposures from 28 August 2018 to 30 September 2019, for which data are available\footnote{\url{https://
chime-frb-open-data.github.io}.}. There are 19 radio bursts detected with CHIME within the $\pm 0.9$-d window around as many peaks\footnote{The number of bursts is equal to the number of peaks by accident.}. To each peak $i$ we assigned the probability $p_i=E_i/E$ for a burst to occur within its $\pm 0.9$-d window, where $E_i$ is the exposure of that window and $E=\sum_i E_i$ is the total exposure. From the multinomial distribution we then calculated the information $I$ of the real sample $\{N_i\}$, where $N_i$ is the number of FRBs observed in window $i$ as follows:
\begin{equation}
I = -\ln{\Big(P_{\rm multi}(\{N_i\})\Big)} = -\ln{(N!)} + \sum_{i}\Big(\ln{(N_i!)} - N_i\,\ln{p_i}\Big)\;,
\end{equation}
where $N = \sum_i N_i = 19$. We then generated $10^5$~samples with $N$ FRBs distributed over the same exposures and compared the distribution of the simulated information content with the real value. As a result, only for the $1.7$\% of simulated samples the information was higher than the real one. In other words, under the assumption of a constant rate around all peaks, the probability of having a distribution equally or less probable than the observed one is $1.7$\%, equivalent to $2.4\,\sigma$ (Gaussian).
In conclusion, although this assumption cannot be rejected with the present data, it suggests that different periods could be characterised by different radio activity at peak. Very recently, the upgraded Giant Metrewave Radio Telescope (uGMRT; \citealt{uGMRT}) detected 15 bursts from \FRB\ in three successive cycles and found extreme variability during the active phase around peak \citep{Marthi20}. Should this be strengthened by future data, we cannot reject the possibility that \FRB\ emitted no FRB during these \HXMT\ observations.

\section{Discussion}
\label{sec:disc}
The discovery of FRB\,200428, a sub-energetic FRB from recently reactivated Galactic source SGR\,1935+2154, provided the first compelling evidence that magnetars are occasionally FRB sources. The question as to whether they are also responsible for the more energetic extragalactic siblings of FRB\,200428 is still open. In particular, some extragalactic FRBs could be due to the most energetic subset of the magnetar population, which in principle could be way more energetic than the so-far known Galactic sources \citep{Margalit20}. A possibility is offered by young (age $\lesssim 10^9$~s) hyperactive magnetars with internal fields $B\sim B_{16}\times 10^{16}$~G and magnetic energy reservoir of $E\sim 2\times 10^{49}\,B_{16}^2$~erg \citep{Beloborodov17}, especially like the ones that can be formed in compact mergers and whose magnetic activity would be enhanced by the large differential rotation at birth \citep{Beloborodov20}. Thus, looking for magnetar burst activity from known extragalactic FRB sources has gained prominence.

The initial spikes of giant flares from Galactic magnetars so far observed have $L\lesssim 10^{47}$~erg/s, $E\lesssim 10^{46}$~erg, and durations $\Delta\,t\lesssim 0.1$~s \citep{Mazets79,Feroci99,Hurley99,Palmer05,Hurley05,Mazets99}. When giant flare candidates of extragalactic origin are considered, the observed released energies and luminosities can be larger by more than one order of magnitude (e.g., \citealt{Mazets08}).
In this context, our upper limits exclude the occurrence of giant flares similar to more energetic than the brightest ones observed from known Galactic magnetars, at least during 23\% of the 3-day window centred on one of the peaks of the expected radio burst activity of \FRB\,. This holds true regardless of the possible simultaneous occurrence of radio bursts.


Hard X-ray bursts as energetic as the one associated with Galactic FRB\,200428 are way below our sensitivity limits and could not be detected in any case. Even assuming the same $\gamma$-to-radio fluence ratio, which lies in the range $5\times10^4$--$3\times10^5$, and rescaling the energy range of \FRB\ radio bursts ($10^{37}$--$4\times10^{38}$~erg), each potentially associated high-energy burst should have $E\sim 3\times10^{42}$--$10^{44}$~erg, which is below our limits.

The fine temporal coincidence between radio and hard X-rays in the case of SGR\,1935+2154 points toward a causal link between the two. That the hard X-ray burst also exhibits some unusual features, like the temporal profile and the spectral hardness, is likely connected with the rarity of the joint manifestation. High-energy bursts are thought to originate in the magnetar magnetosphere as a result of some twisting and buildup of free magnetic energy, which is suddenly released through reconnection and consequent pair plasma acceleration. These processes could either be induced by crustal fractures or have a magnetospheric origin \citep{Lyutikov03b}, where the former seems to be favoured in the case of GFs \citep{Feroci01,Hurley05}.
In this context, FRBs could be coherent curvature radiation from pairs  \citep{Katz14,Kumar17,YangZhang19} or due to an yet-unidentified process  \citep{LyutikovPopov20}, but in any case, taking place in the magnetosphere, a scenario which would account for the simultaneity of radio and hard X-rays \citep{LiHXMT20}. Alternatively, FRBs could be synchrotron maser radiation caused by relativistic magnetised shocks driven by plasmoids at outer radii ($10^{14}$--$10^{16}$~cm), that were launched by flares \citep{Lyubarsky14,Beloborodov17,Beloborodov20,Metzger19}.
In this scenario FRBs would be much more collimated than the high-energy emission due to relativistic beaming of the plasmoids: this would explain both the negligible time delay between radio and hard X--rays, and the rarity of the FRB emission associated with magnetar bursts. Besides, the energy ratio between flare and the associated FRB, $\sim 10^5$, is compatible with the expected radiative efficiency \citep{PlotnikovSironi19,Beloborodov20}.

While FRB\,200428 could belong to the low-energy tail of the extragalactic FRB energy distribution, which cannot be explored at cosmological distances with the current instrumentation \citep{Bochenek20}, the question as to whether the same mechanism at play for SGR\,1935+2154 can be scaled up by 4--6 decades in the case of GFs, remains open and therefore justifies the search for them from FRB sources.

The origin of the periodicity found in the radio activity of \FRB, which might also be the case for FRB\,121102 \citep{Rajwade20}, and its implications on the possible high-energy flaring activity of the putative magnetar, depend on the model.
Although the possibility that the periodicity is due to the star's rotation period is not ruled out \citep{Beniamini20}, one can identify two main alternative scenarios: 1) a tight binary system, where a classical magnetar emission is modulated by the phase-dependent absorption conditions related to the massive companion's wind \citep{Lyutikov20}, or where the modulation reflects the orbit-induced spin precession \citep{YangZou20}, or other variants \citep{Gu20,IokaZhang20}; 2) an isolated precessing magnetar \citep{Levin20,ZanazziLai20,Sobyanin20}, where, among the different possibilities, Lens-Thirring precession due to a tilted disc could be an option \citep{Chen20}.

In the context of a systematic monitoring of the high-energy activity of \FRB, our observations help constrain the rate of possible GFs with a unique broadband sensitivity from 1 to 100~keV, with upper limits on the released energy of possible bursts of $E\lesssim10^{46}$~erg or even less for durations $\Delta\,t\lesssim0.1$~s. Considering the different energy bands, these values are comparable with those obtained in the X--rays with {\em Chandra} and {\em XMM-Newton} \citep{Scholz20,Pilia20} and significantly more sensitive than those obtained with all-sky monitors such as {\em Fermi}/GBM \citep{Scholz20}. Under the assumption that the frequency of radio bursts of \FRB\ around peak is the same for all periods, our observations almost certainly covered one or more of them. However, the analysis of the CHIME data suggests that different cycles could be characterised by different radio activity around peak.

\section{Conclusions}
\label{sec:conc}
We followed up the periodic FRB repeating source \FRB, which also happens to be the closest extragalactic FRB source to date ($149$~Mpc), during the peak expected from February 4 to 7, 2020, with the three instruments aboard \HXMT, exploiting its unique combination of sensitivity and broadband. We searched for burst activity with a duty cycle of $\sim 1/4$ and found nothing down to $E\lesssim10^{46}$~erg (1--100~keV energy band) or even less for durations $\Delta\,t\lesssim0.1$~s. This rules out the occurrence of the most energetic giant flares yet observed from Galactic magnetars. No other observations of \FRB\ and, especially, no radio burst around that peak have been reported to date.

Assuming that its radio burst activity around peak is the same for all periods, our observations almost certainly covered some bursts. Nevertheless, presently available CHIME data suggest that the source likely experiences different degrees of radio activity at peak across different cycles, leaving the possibility that \HXMT\ monitored the source during a burst-free interval.

Yet, the search for magnetar flaring activity is motivated by two main reasons: 1) at least a sizeable fraction of extragalactic FRB sources are likely to be magnetars, that are possibly more active and have stronger magnetic fields than their Galactic siblings; 2) the complex relation between FRBs and simultaneous hard X-ray bursts revealed by SGR\,1935+2154 is still to be understood. Only through systematic multi-wavelength campaigns will the nature and the role of extragalactic magnetars as FRB sources be clarified. In this respect, the \HXMT\ observations reported in the present work also served as a test bed and calibration of the search methods expressly devised for this purpose, with regard to the future joint campaigns that are planned for \FRB\ as well as for other suitable repeating FRBs.

\begin{acknowledgements}
We thank the referee Kevin Hurley for the swift and detailed comments that improved the manuscript. This work is supported by the National Program on Key Research and Development Project (2016YFA0400800) and the National Natural Science Foundation of China under grants 11733009, U1838201 and U1838202. This work made use of data from the {\em Insight}-HXMT mission, a project funded by China National Space Administration (CNSA) and the Chinese Academy of Sciences (CAS).
We acknowledge financial contribution from the agreement ASI-INAF n.2017-14-H.0.
We acknowledge use of the CHIME/FRB Public Database, provided at \url{https://www.chime-frb.ca/} by the CHIME/FRB Collaboration.
\end{acknowledgements}

\appendix

\section{Study of background properties}
\label{sec:app_A}
\begin{figure*}
\centering
\includegraphics[width=\linewidth]{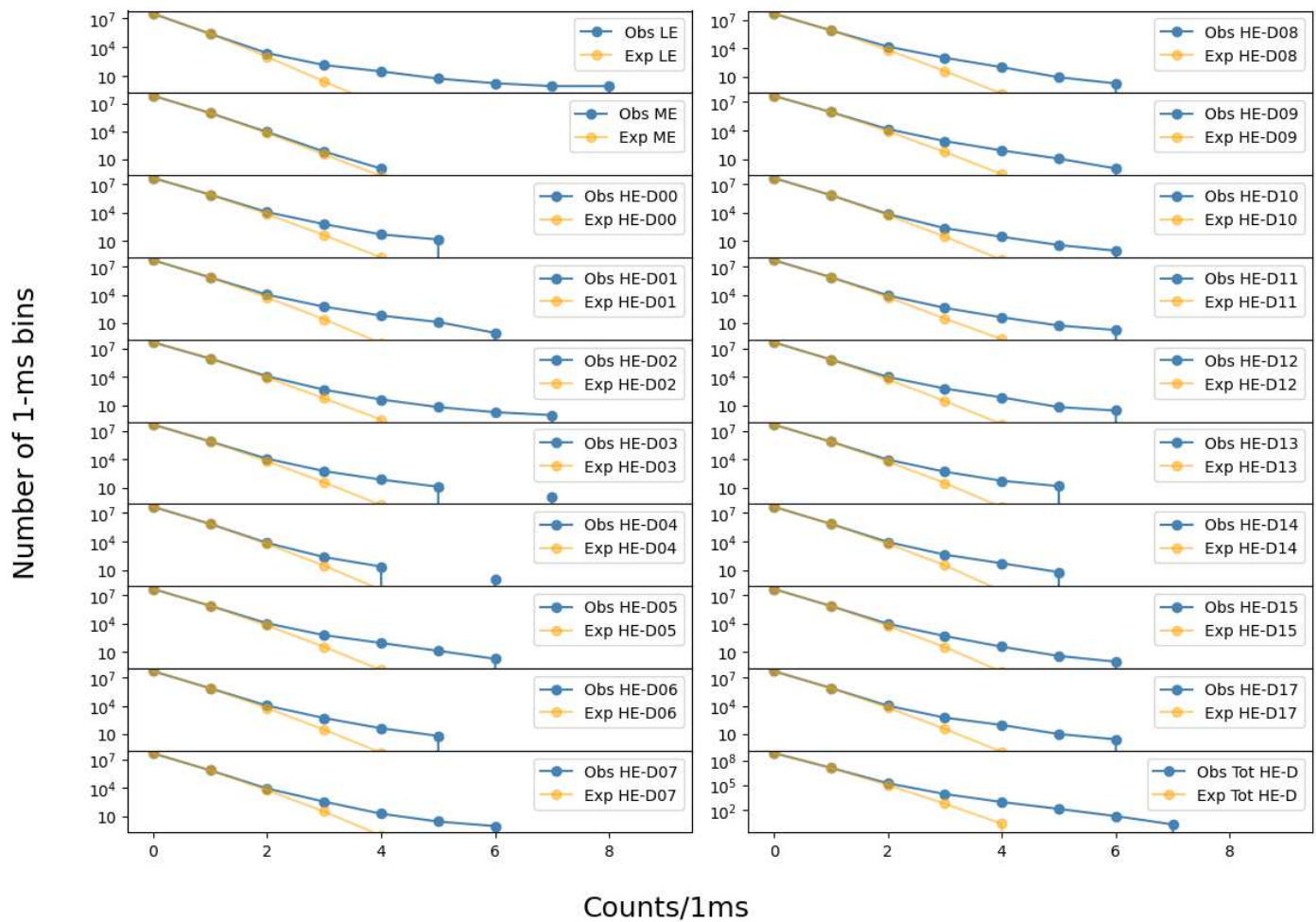}
\caption{Total 1-ms count distribution for each of the 19 detectors (LE, ME, and 17 HE-NaI units): blue solid lines show the observed distribution, while the orange ones show the same distributions expected from a varying background model purely affected by Poisson noise. The bottom right panel shows the summed distribution of the HE units.}
\label{fig:cts_distrib_all}
\end{figure*}

The study of the statistical noise affecting the background time series of each of the 19 detectors (LE, ME, and 17 HE-NaI units) is important to establish the confidence level of any possible transient candidate. In order to characterise it, for each detector we first derived the overall 1-ms count distribution. As already mentioned in Section~\ref{sec:statnoise}, we then compared each observed count distribution with the corresponding synthetic one, obtained from statistical realisations of a variable Poisson process, whose expected value, as a function of time bin, is given by the local background, that estimated with the procedure described in Section~\ref{sec:dataset}. The resulting distributions are shown together in Figure~\ref{fig:cts_distrib_all} with semi-logarithmic scale. The blue and yellow histograms correspond to the real and the Poisson-expected count distributions, respectively. Clearly, in all cases there is evidence for a small component in excess of the pure Poisson noise case, which is responsible for a higher-than-expected number of 1-ms bins with $\gtrsim 4$~counts. More precisely, the extra-Poissonian variance affecting the real counts amounts to $1.2$\%, $0.6$\%, and $0.5$\% for the LE, ME, and average HE, respectively. Following the detailed inspection of these excesses, each of them is found in one individual detector at a time, so they are completely uncorrelated among different detectors and, as such, they are incompatible with what is expected from a plane wave. These spikes are therefore spurious and are likely to be mainly due to cosmic rays of high atomic number Z. They excite metastable states in the crystals, giving rise to a short-lived phosphorescence, which is nonetheless long enough to have the electronics detect several counts. This is probably the same effect as the one observed in the {\em BeppoSAX} Gamma--Ray Burst Monitor during the first months, before the lower-energy thresholds were finally increased \citep{Feroci97}. 

%

We tested this possibility in the case of HE data, by selecting all the 1-ms bins with $\ge 6$~counts and studying the distribution of these counts as a function of energy, considering three energy channels: $25$--$30$, $30$--$40$, and $40$--$80$~keV. As a result, 80\% of them lie below $30$~keV, while none of them passes the $40$-keV threshold.

\begin{figure}
\centering
\includegraphics[width=0.5\textwidth]{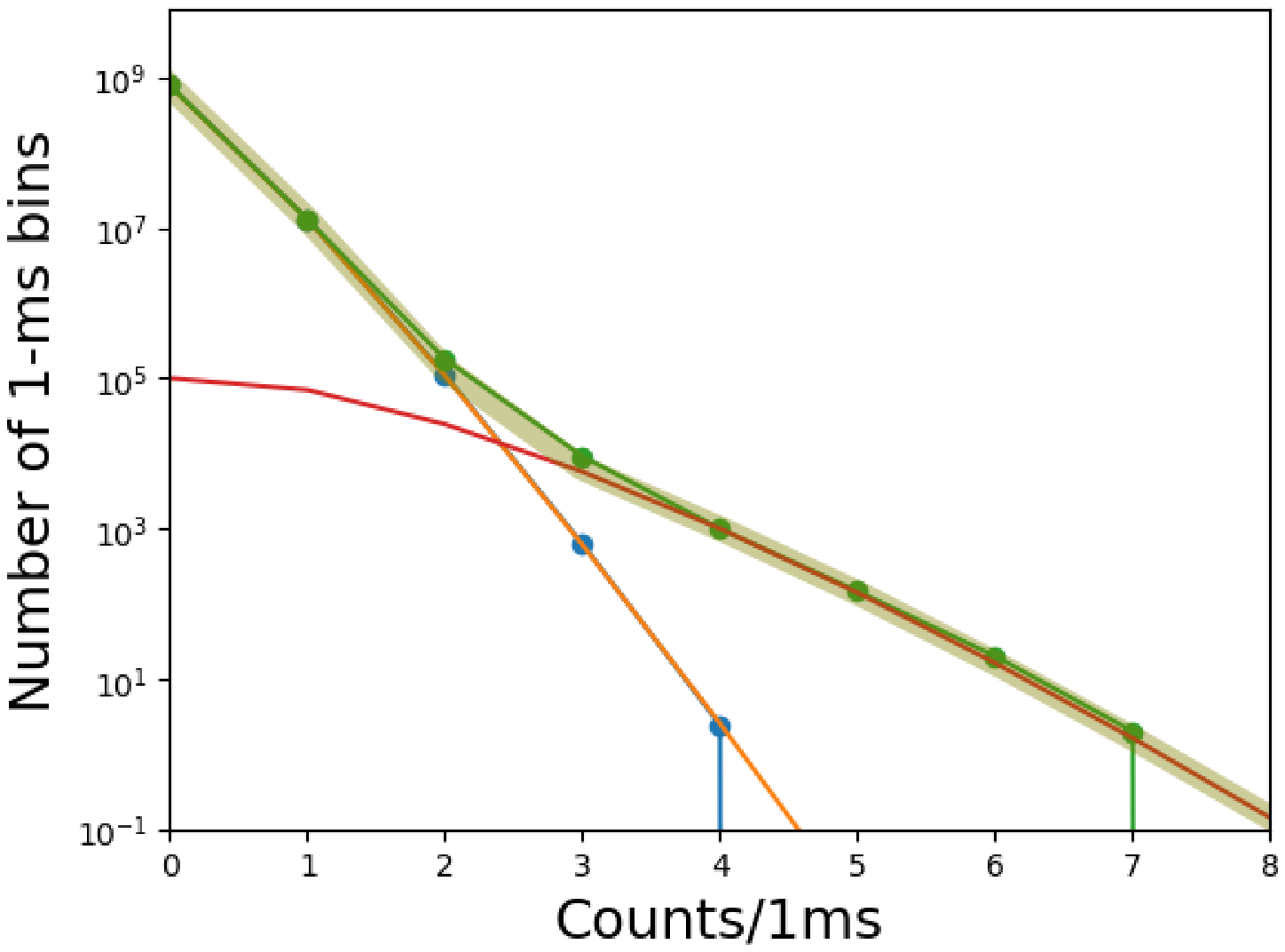}
\caption{Total 1-ms count distribution of the 17 HE-NaI units (green points). It is modelled (green thick line) as the sum of two independent components: in addition to the one that accounts for $\sim 99$\% of the total counts with an average count rate of $0.017$~counts\,ms$^{-1}$, (blue points and orange solid line) there is an additional component (red solid line), which is responsible for the remaining $1$\% of counts, with an average count rate of $0.7$~counts\,ms$^{-1}$ and whose exposure is just $2.6\times10^{-4}$ times the total net one.}
\label{fig:cts_distrib_HE}
\end{figure}

We further characterised the extra-Poissonian component in the HE case by modelling the observed total 1-ms count distribution, that is the summed distribution of all 17 HE-NaI detectors, shown in the bottom right panel of Fig.~\ref{fig:cts_distrib_all}. We found out that the overall distribution can reasonably be modelled as the sum of two independent Poisson processes with different constant rates. The dominant one has an average rate of $0.017$~counts~ms$^{-1}$, while the other has a much larger rate ($0.7$~counts~ms$^{-1}$), but is active only for a limited amount of time, whose total exposure is just $2.6\times10^{-4}$ times that of the dominant one. The result is shown in Figure~\ref{fig:cts_distrib_HE}. Therefore, one may look at the extra-Poissonian component as due to the particle spikes that occasionally impact the detectors for a very limited fraction of the total observing time.

%

\section{Trigger criteria of the Multi-Detector Search}
\label{sec:app_B}
The MDS conceived for the available data set of three instruments consists of three alternative criteria. Whenever the counts for a given time bin fulfil at least one of the three criteria, the corresponding event is promoted to the status of transient candidate. They mainly differ in the combinations of data sets to which they can be applied. Each of them is defined through a set of thresholds, expressed in Gaussian $\sigma$ units as in \citet{Guidorzi20}, that must be exceeded by the counts recorded in a given time bin. These thresholds depend on the integration time, on the kind of detector, as well as on the number of detectors with available data.
Table~\ref{tab:trigger} reports all of them. All the parameters were tuned so as to end up with a relative small number of false positives ($\lesssim 1$) taking into account the multi-trials due to the total number of screened bins. As explained in Section~\ref{sec:MDS}, we started with a initial set based on the pure Poisson noise assumption. Next, in order to account for the impact of a small, but significant extra-Poissonian component due to the presence of occasional particle spikes, we refined them by means of Monte Carlo simulations. These were carried out by shuffling the observed data sets 100 times, so preserving the properties of the count distributions of each individual detector. We then applied the MDS to these 100 statistically equivalent data sets to evaluate the false positive rate.

While the number of bins to be screened obviously decreases with increasing integration time, looking at Table~\ref{tab:trigger} one might be puzzled by the way the minimum number of HE units varies with integration time. These values were obtained taking into account the granularity of counts as a discrete and not a continuous distribution. This has an impact especially at the shortest integration times, for which the expected background counts per bin are $\ll 1$. As a consequence, raising the threshold on counts  by just one unit, implies a drastic change in the corresponding probability (see G20 for further explanations).
\begin{table*}
\caption{Trigger criteria of the MDS.}
\label{tab:trigger}
\centering\begin{tabular}{|c|l|r|c|c|c|c|}
\hline\hline
Criterion & Combinations of & $\Delta\,t^{\rm (a)}$ & $n_{\rm LE}^{\rm (b)}$ & $n_{\rm ME}^{\rm (b)}$ & $n_{\rm HE}^{\rm (b)}$ & $N_{\rm min,HE}^{\rm (c)}$\\
& data sets & (ms) & & & & \\
\hline
\multirow{6}{*}{1} & \multirow{3}{*}{LE \& HE} &    1 & $2.7$ & $2.7$ & $1.8$ & 7\\ 
 &         &    4 & $2.7$ & $2.7$ & $1.8$ & 8\\ 
 &  &   10 & $2.5$ & $2.5$ & $1.8$ & 6\\ 
 & \multirow{3}{*}{ME \& HE}      &   64 & $2.4$ & $2.4$ & $1.8$ & 7\\ 
 &  &  256 & $2.4$ & $2.4$ & $1.8$ & 7\\ 
 &       & 1024 & $2.3$ & $2.3$ & $1.8$ & 7\\
 \hline
 \multirow{6}{*}{2} & \multirow{6}{*}{LE \& ME} &    1 & $3.5$ & $3.5$ & - & -\\
 &        &    4 & $3.5$ & $3.5$ & - & -\\
 &      &   10 & $3.5$ & $3.5$ & - & -\\
 &       &   64 & $3.5$ & $3.5$ & - & -\\
 &       &  256 & $3.5$ & $3.5$ & - & -\\
 &       & 1024 & $3.5$ & $3.5$ & - & -\\
\hline
\multirow{6}{*}{3} & \multirow{6}{*}{HE alone} &    1 & -     &  -    & $1.8$ & 8\\
&         &    4 & -     &  -    & $1.8$ & 10\\
&        &   10 & -     &  -    & $1.8$ & 7\\
&        &   64 & -     &  -    & $1.8$ & 8\\
&        &  256 & -     &  -    & $1.8$ & 8\\
&        & 1024 & -     &  -    & $1.8$ & 8\\
\hline
\end{tabular}
\begin{list}{}{}
\item[$^{\rm (a)}$]{Integration time of a single bin.}
\item[$^{\rm (b)}$]{Threshold on individual bin, expressed in Gaussian $\sigma$ units, for each detector unit.}
\item[$^{\rm (c)}$]{Minimum number of HE units to be triggered.}
\end{list}
\end{table*}

The three criteria are explained in more detail as follows:
\begin{enumerate}
    \item This criterion demands that at least two instruments, one of which must be HE, fulfil the corresponding trigger condition. Given that HE consists of 17 independent units, the HE trigger condition includes a minimum number of HE units for which the corresponding threshold must be exceeded.
    \item In this case, the combination of LE and ME is considered, regardless of HE. Compared with criterion 1, because of the lack of the multiple detector of different HE units, one must increase the thresholds on the LE and ME to avoid an excessively high rate of statistical flukes that may trigger it.
    \item This criterion concerns HE data, regardless of the two softer instruments. Compared with criterion 1, the minimum number of HE units to be triggered is slightly higher, to compensate the lack of information from the other two instruments.
\end{enumerate}
There are a couple of important points to make: the time bins for which data from all instruments are available, are screened through all three criteria. Secondly, the criteria are not strictly mutually exclusive, especially for bright events. For instance, a bright transient, for which filtered data from the three detectors are available, would trigger criterion 1 with all instruments, but could also trigger the other two criteria as well. Conversely, regardless of the event brightness, whenever for some reasons only filtered data from two instruments (or from HE alone) are available -- and this is indeed the case for a non-negligible fraction of net exposure as reported in Table~\ref{tab:log}--, criteria 2 and 3 come into play and make sure that the transient is not missed by the trigger logic. Furthermore there are other possible cases, for which the spectral properties of the event makes it detectable only through specific criteria: when this is particularly soft, criterion 2 is more likely to trigger, whilst criterion 3 could work best when it is relatively dim and hard.

That the three criteria are not mutually exclusive, demands that the expected false positive rate should be better estimated by applying them to simulated data sets, that preserve the statistical properties of the count distributions of the individual detectors. This is reason why we opted for this choice. The results are reported in Table~\ref{tab:FPrate}: the number of expected false positives already accounts for the total number of screened time bins and includes all the three trigger criteria together.




\end{document}